\documentclass[aps,pra,10pt,twocolumn,showpacs]{revtex4-1} 
\usepackage{amsfonts,amsmath,amssymb}
\usepackage{graphicx}
\usepackage{microtype}
\usepackage[utf8]{inputenc}
\usepackage[pdfstartview=FitH,colorlinks=true,linkcolor=blue,citecolor=blue,urlcolor=blue]{hyperref}
\usepackage[font=small,format=hang,labelfont={sf,bf}]{caption}

\begin{document}

\title{Collective oscillations of a trapped quantum gas in low dimensions}

\author{Giulia De Rosi}
\email{giulia.derosi@unitn.it}

\author{Sandro Stringari}
\email{stringar@science.unitn.it}

\affiliation{INO-CNR BEC Center and Dipartimento di Fisica, Universit\`a di Trento, Via Sommarive 14, I-38123 Povo, Italy}

\date{\today}

\begin{abstract}
We present a comprehensive study of the discretized modes of an atomic  gas in different conditions of confinement. Starting from  the equations of hydrodynamics we derive a closed equation for the velocity field, depending on the adiabatic and isothermal compressibilities and applicable to different dimensions and quantum statistics. At zero temperature the equation reproduces the irrotational behavior of superfluid hydrodynamics. It is also applicable above the critical temperature in the collisional regime, where the appearence of rotational components in the velocity field is caused by the external potential. In the presence of harmonic trapping, a general class of analytic solutions is obtained  for systems exhibiting a polytropic equation of state, characterized by  a power law isoentropic dependence of the pressure on the density. Explicit results for the compressional modes are derived for both Bose and Fermi gases in the pancake, cigar as well as in the deep 2D and 1D regimes. Our results agree with the analytical predictions  available in the literature in some limiting cases. They are particularly relevant  in 1D configurations, where the study of the collective frequencies  could provide a unique test of the achievement of the collisional regime at finite temperature. 
\end{abstract}

\pacs{PACS numbers}

\maketitle

\section{Introduction}
\label{Sec:Introduction}

In recent years the study of low-dimensional quantum gases has been the object of  significant experimental and theoretical research \cite{Pitaevskii2003, Pricoupenko2004}. The main interest is due to the  increase of quantum correlations and fluctuactions caused by the reduced dimensionality which make the properties of these systems significantly different from their 3D counterpart \cite{Bloch2008}.

In two dimensions, thermal fluctuations rule out the existence of long range order  at finite temperature (Hohenberg-Mermin-Wagner theorem) \cite{Mermin1966, Hohenberg1967} and novel aspects of superfluid phenomena, like the Berezinskii-Kosterlitz-Thouless transition,  take place. In compressible 1D systems the occurrence of long range order is ruled out by quantum fluctuations even at  $T=0$ \cite{Pitaevskii1991, Pitaevskii2003}. 
The goal of this paper is to focus on the collective oscillations of a low dimensional atomic gas. While some of the results discussed in this paper have been already derived in the literature, a  unified description covering both Bose and Fermi statistics at zero as well as at finite temperature in different regimes of axial and radial  trapping is still missing. Our  analysis can help in providing useful links and comparisons among different regimes of experimental and theoretical interest.

Our investigation is based on the hydrodynamic formalism applicable to various regimes, including the superfluid regime at zero temperature and the collisional regime in the non superfluid phase  above the critical temperature. To this purpose, we will derive a general formulation of the hydrodynamic equations in presence of an external trapping potential in terms of the velocity field. This approach  explicitly points out the irrotational vs rotational nature of the solutions and the role of the equation of state. It reduces to a simplified form when the equation of state exhibits a polytropic dependence on the density, allowing for an important class of analytic solutions. Our results are also compared with the predictions for the  frequencies of the normal modes in the collisionless regime.

The paper is organized as follows: 

In Sec. \ref{Sec:polytropic equation} we introduce the concept of polytropic equation of state, both in terms of the pressure and, at $T=0$, of the chemical potential. The values of the polytropic coefficient $q$, fixing the power law isoentropic dependence of the equation of state on the density, are discussed in some relevant  cases of different dimensionality  for both Bose and Fermi gases at zero as well as at finite temperature. We will consider geometrical configurations corresponding to  pancake and cigar profiles, whose equation of state can be derived starting from the 3D equation of state,  employing the local density approximation along the  directions of the confinement. We will also discuss the  case of  deep 2D and 1D configurations where the motion of the gas is instead frozen along the directions of the confinement.

In Sec. \ref{Sec:hydrodynamics} we  derive a general  equation obeyed by  the velocity field associated with the collective motion,  starting from hydrodynamic theory in the presence of external trapping. This equation holds both for a superfluid gas at zero temperature, where the macroscopic dynamic behavior  is described by the hydrodynamic theory of superfluids,  and in the normal phase,   in the collisional regime, characterized by the condition $\omega\tau \ll 1$, where $\tau$ is a typical  collisional time.  We will show that  the equation for the velocity field acquires a particular simple form if the equation of state of the gas exhibits a polytropic dependence on the density.

In Secs. \ref{Sec:surface}, \ref{Sec:2D_isotropic}, \ref{Sec:coupled} and \ref{Sec:1D} we derive the discretized oscillation frequencies in the presence of additional harmonic trapping in the plane (in the case of 2D configurations) and along the axis (in the case of 1D configurations).  Special emphasis is given to the lowest breathing modes  whose frequency, differently from the divergency free solutions,  is explicitly  sensitive to the value of the polytropic coefficient and  can exhibit a  temperature dependence even in the hydrodynamic regime.  

\section{Polytropic equation of state}
\label{Sec:polytropic equation}

A useful feature, which allows for a significant simplification of the theoretical analysis, is the fact that in several  configurations, corresponding to  different dimensional,  interaction and temperature regimes and quantum  statistics, the equation of state of the gas exhibits a simple power law dependence on the density $n$ in the form 
\begin{equation}
P(n,\bar{s})=n^q p(\bar{s}) \; ,
\label{Eq:q}
\end{equation}
where $P$ is the pressure of the gas, $q$ is  called the polytropic coefficient, $\bar{s}$ is the entropy per particle  and $p(\bar{s})$ is a function of  $\bar{s}$, fixed by the thermodynamic behavior of the gas at zero ($\bar{s}=0$) as well as at finite temperature. 
As we will show, the polytropic coefficient $q$ plays a crucial role in determining the frequency of the discretized collective oscillations in the presence of harmonic trapping. Viceversa, the value of the function $p(\bar{s})$ determines the adiabatic sound velocity, according to the thermodynamic law
\begin{equation}
mc^2= \left(\frac{\partial P}{\partial n}\right)_{\bar{s}} = qn^{q-1} p(\bar{s}) = q \frac{P}{n} \; .
\label{c2}
\end{equation}

Starting from Eq. \eqref{Eq:q} and using the thermodynamic relation $\left(\partial U/\partial n\right)_{\bar{s}} = P/n^2$, one can write the following  relationship between the energy per particle and the pressure of the gas:
\begin{equation}
\label{Eq:U}
U = \frac{1}{q-1}\frac{P}{n}
\end{equation}

At $T=0$ the scaling law (\ref{Eq:q}) can be also written in the form
\begin{equation}
\mu(n,T=0) = \frac{q}{q-1}n^{q-1} p(\bar{s}=0)
\label{Eq:mu}
\end{equation}
where $\mu$ is the chemical potential of the gas related to the pressure by the Gibbs-Duhem thermodynamic relation
\begin{equation}
dP = nd\mu +sdT
\label{GD}
\end{equation}
where $s=n\bar{s}$ is the entropy density. 
The  sound velocity at zero temperature, where $dP=nd\mu$,   can be rewritten in terms of the chemical potential as 
\begin{equation}
mc^2 = n\left(\frac{\partial \mu}{\partial n}\right)_{\bar{s}} = (q - 1)\mu \; .
\label{Eq:c2_T0}
\end{equation}

When applied to  configurations of lower dimensions the pressure and the density entering (\ref{Eq:q}) should be replaced by  the corresponding 2D or 1D pressure and density of the gas.

The parametrization (\ref{Eq:q}) for the equation of state applies to an  important class of physical systems. The values of the corresponding polytropic index $q$ are explicitly derived in Appendix \ref{Sec:EOS} and are reported in the Tables \ref{Tab:qTF} and \ref{Tab:qdeep}. For example, in a   3D  Fermi gas at unitarity \cite{Giorgini2008} (as well as in the ideal 3D Fermi gas and in the ideal 3D Bose gas above the critical temperature) simple dimensionality arguments permit to identify the value $q= 5/3$ for the polytropic coefficient. The 3D weakly interacting Bose gas \cite{Dalfovo1999}, at zero temperature, is instead characterized by the different value $q=2$.  On the other hand, Eq. \eqref{Eq:q} does not hold for the ideal Bose gas below the condensation temperature, since this system has an infinite compressibility and its pressure vanishes in the limit of zero temperature. 

In the presence of axial or radial confinement one should distinguish between the case when the 3D equation of state can be still applied locally within the local density approximation (LDA) and the case where the motion is instead frozen along the directions of confinement. 

The first case includes the so called pancake geometry (where the equation of state can be expressed in terms of the 2D density and pressure) and the so called cigar geometry (where the equation of state can be expressed in terms of the 1D density and pressure). Both configurations have been realized experimentally both for bosons \cite{Jin1996, Mewes1996} as well as for fermions \cite{Weimer2015, Kinast2004, Bartenstein2004} and are characterized by different values of the polytropic coefficient (see Table \ref{Tab:qTF}).
The second case includes the deep 2D and 1D configurations where the thermodynamic behavior cannot be derived starting from the 3D equation of state in the LDA. These configurations are particularly interesting from the many-body and thermodynamic point of view. In 2D they exhibit the Berezinski-Kosterlitz-Thouless transition \cite{Desbuquois2012}, while in the 1D case for bosons, at zero temperature, they are described by Lieb-Lininger theory \cite{Lieb1963}. Some of these low dimensional regimes are also characterized by well defined values of the polytropic coefficient which are reported in Table \ref{Tab:qdeep}.

It is finally  interesting to compare the adiabatic sound velocity $c_{3D}$ of 3D uniform systems  with the sound velocity $c$ for the lower dimensional regimes   reported in Table \ref{Tab:qTF}. At $T=0$, for equal values of the chemical potential (which, in the case of the LDA regimes,  would correspond to equal values of the central density),  one finds the following relationship 
\begin{equation}
\label{Eq:cLDA_zeroT}
\frac{c^2}{c_{3D}^2} = \frac{q - 1}{q_{3D} - 1}\; .
\end{equation}
where $q_{3D}$ is the polytropic coefficient of the equation of state of 3D uniform matter. In the special case of a cigar trap (see Table \ref{Tab:qTF}), the above equation gives the value $c^2/c_{3D}^2 = 1/2$ for a weakly interacting  Bose gas and  $c^2/c_{3D}^2 = 3/5$ for the unitary Fermi gas. The above reduction factors  of the sound velocity were first theoretically derived by Zaremba  \cite{Zaremba1998} and by  Capuzzi et al. \cite{Capuzzi2006}. They were confirmed experimentally by Andrews et al  \cite{Andrews1997} and Joseph et al. \cite{Joseph2007} for cigar Bose and Fermi gases, respectively.
In the classical regime of high temperature, where $P=nkT$, one instead find the result 
\begin{equation}
\label{Eq:cLDA_Tcl}
\frac{c^2}{c_{3D}^2} = \frac{q}{q_{3D}}\; ,
\end{equation}
where the sound velocities are calculated at the same temperature.  

\begin{table}
\caption{Polytropic index $q$ for the weakly interacting Bose gas at $T=0$ and for the unitary Fermi gas in different LDA regimes. The results for the unitary Fermi gas hold also for the ideal Fermi gas at all temperatures and for the ideal  Bose gas above $T_c$.
 }
\label{Tab:qTF}
\begin{tabular}{ccc}
 \hline\noalign{\smallskip}
 \hline\noalign{\smallskip}
&\textbf{Bose gas (T = 0)}&\textbf{Unitary Fermi gas} \\
\noalign{\smallskip}\hline\noalign{\smallskip}
\textbf{3D uniform} & $2$ & $5/3$    \\
\noalign{\smallskip}\hline\noalign{\smallskip}
\textbf{pancake (LDA)} &   $5/3$  &   $3/2$   \\
\noalign{\smallskip}\hline\noalign{\smallskip}
\textbf{cigar (LDA)} &   $3/2$  &   $7/5$   \\
\noalign{\smallskip}\hline 
\noalign{\smallskip}\hline 
\end{tabular}
 \end{table}
\begin{table}
\caption{Polytropic index $q$ for a Bose gas for different 2D and 1D regimes in the presence of tight confinement. The 2D mean field value $q = 2$ holds also for the interacting Fermi gas in the BCS and BEC limits.}
\label{Tab:qdeep}
\begin{tabular}{ccc}
 \hline\noalign{\smallskip}
 \hline\noalign{\smallskip}
&\textbf{T = 0} & \textbf{high T} \\
\noalign{\smallskip}\hline\noalign{\smallskip}
\textbf{2D mean field} & $2$ & $2$    \\
\noalign{\smallskip}\hline\noalign{\smallskip}
\textbf{1D mean field} &   $2$  &   $3$   \\
\noalign{\smallskip}\hline\noalign{\smallskip}
\textbf{1D Tonks-Girardeau} &   $3$  &   $3$   \\
\noalign{\smallskip}\hline 
\noalign{\smallskip}\hline 
\end{tabular}
 \end{table}


\section{Hydrodynamic Equations in  the presence of external trapping}
\label{Sec:hydrodynamics}

In this Section we discuss the  hydrodynamic (HD) equations for the velocity field  describing the collective motion of the gas in the presence of an external confining potential and in the dissipationless (entropy conserving) regime. Even if these equations have a   classical shape, quantum mechanical effects deeply affect their solutions, being responsible for important changes in the behavior of the equation of state and, in particular, in the value of the polytropic coefficients, as discussed in the previous Section. 
These equations describe correctly the dynamic behavior of an interacting  system at zero temperature, where they coincide with the irrotational hydrodynamic equations of superfluids and apply to interacting  Bose and Fermi superfluid gases, as well as to strongly interacting superfluids like $^4He$ \cite{Pitaevskii2003}. They also apply above the critical temperature $T_C$ for superfluidity, in the collisional regime where $\omega \tau \ll 1$, with $\tau$ the collisional time, and the equation of state can still depend in a crucial way on quantum statistical effects \cite{Griffin1997}. For the ideal Fermi gas they were derived in the collisional limit in \cite{Minguzzi2001} at T = 0. For finite temperature, below $T_C$, the hydrodynamic equations should be instead generalized to include the coupled  description of the normal and superfluid components (Landau's two-fluid hydrodynamic equations), allowing for the propagation of first and second sound (see, for example, \cite{Pitaevskii2003}). In the presence of harmonic trapping the  hydrodynamic equations have proven quite successful at zero temperature in predicting the frequencies of the collective oscillations \cite{Stringari1996}, in fair agreement with the experimental findings.

The hydrodynamic equations include the equation of continuity 
\begin{equation}
\label{Eq:continuity2} 
\frac{\partial n(\bold{r},t)}{\partial t} + \bold{ \nabla} \cdot \left[n_0({\bf r})\bold{v}(\bold{r},t) \right] = 0, 
\end{equation}
 the equation for the entropy density
\begin{equation}
\label{Eq:continuity_entropy}
\frac{\partial  s({\bf r },t)}{\partial t} = - \nabla \left[s_{0}({\bf r}) {\bf v}({\bf r}, t)\right],
\end{equation}
and the Euler equation for the current density ${\bf j}({\bf r}, t)=m n_0({\bf r}){\bf v}({\bf r}, t)$
\begin{equation}
\frac{\partial {\bf j}({\bf r},t)}{\partial t} = - \left[\nabla P({\bf r}, t) + n({\bf r}, t) \nabla V_{ext}({\bf r})\right]
\label{Eq:j}
\end{equation}
In the above equations $n_0$ and $s_0$ are the equilibrium values of the density and of the entropy density, respectively and, for simplicity, we have considered the limit of small amplitude oscillations and small velocities. At zero temperature, where the equation for the entropy identically vanishes,  the hydrodynamic equations reduce to two coupled equations for the density and the velocity field. 

Starting from the above equations and using suitable thermodynamic relations, it is possible to derive the following equation for the  stationary solutions of the velocity field characterized by the time dependence ${\bf v}({\bf r},t)={\bf v}({\bf r})e^{-i\omega t}$ (see Appendix \ref{Sec:appendix_wave}):
\begin{multline}
m\omega^2 {\bf v}= -\nabla \left[\left(\frac{\partial P}{\partial n}\right)_{\bar{s}} (\nabla \cdot {\bf v})\right] + (\gamma - 1) (\nabla V_{ext})(\nabla \cdot {\bf v})  \\
+ \nabla \left({\bf v} \cdot \nabla V_{ext}\right) 
\label{Eq:eqvideal}
\end{multline}
where the adiabatic coefficient
\begin{equation}
\label{Eq:gamma}
\gamma = \left(\frac{\partial P}{\partial n} \right)_{\bar{s}}/\left(\frac{\partial P}{\partial n} \right)_{T}
\end{equation}
provides the ratio between the isothermal and the  adiabatic compressibilities.
Equation \eqref{Eq:eqvideal} holds for   configurations of different dimensionality and for arbitrary external potentials compatible with the applicability of the local density approximation along the direction of the velocity flow. It shows that the eigenfrequencies $\omega$ of the collective oscillations  can be determined once the adiabatic and the isothermal compressibilities are known. These quantities depend on the nature of the system, on temperature and, of course, on the dimensionality of the configuration.  Result \eqref{Eq:eqvideal} shows explicitly the emergence of rotational components in the velocity field ${\bf v}$ caused by the presence of the external potential (term proportional to $(\gamma-1)$). 
This  term exactly vanishes if the adiabatic and isothermal compressibilities  coincide ($\gamma=1$),  a condition ensured only at $T=0$. In this limit Eq. \eqref{Eq:eqvideal} actually coincides with the hydrodynamic equation of irrotational superfluids.
To our knowledge,  the equations of linearized hydrodynamics, in the presence of an external potential, have never been derived so far in the general form (\ref{Eq:eqvideal}). Their derivation represents one of the main results of the present work. 

Starting from the  equations of hydrodynamics it is also possible to derive (see Appendix \ref{Sec:appendix_wave}) the equation  
\begin{equation}
\label{Eq:T_t}
\frac{\partial T({\bf r}, t)}{\partial t} = - \frac{T}{c_v}\left(\frac{\partial P}{\partial T}\right)_n \frac{\nabla \cdot {\bf v({\bf r}, t)}}{n_0} \; ,
\end{equation}
for the time dependence of the temperature of the gas,  
where $c_v$ is the specific heat at constant volume.
Eq. \eqref{Eq:T_t} shows that, for divergency free solutions ($\nabla \cdot {\bf v} = 0$), like the surface modes (Sec. \ref{Sec:surface}), the temperature of the trapped gas is constant during the oscillation. 

For systems obeying the polytropic behavior (\ref{Eq:q}) the equation for the temperature takes the simplified form
\begin{equation}
\label{Eq:T_t_scaling}
\frac{\partial T({\bf r}, t)}{\partial t} = -(q-1) T \nabla \cdot {\bf v}({\bf r}, t)
\end{equation}
which follows directly from Eq. \eqref{Eq:U} and from the thermodynamic relation $c_v = \left(\partial U/\partial T\right)_n$.
Equation \eqref{Eq:T_t_scaling} coincides with the result derived in \cite{Griffin1997} for a 3D ideal Bose gas at $T \ge T_c$, where $q=5/3$.

For systems obeying the polytropic law (\ref{Eq:q}), the equation for the velocity field \eqref{Eq:eqvideal} takes the further  simplified expression (see Appendix \ref{Sec:appendix_wave}) 
\begin{multline}
\label{Eq:eqvfinal}
m\omega^2{\bf v} = - \left(\frac{\partial P}{\partial n} \right)_{\bar{s}}\nabla \left(\nabla \cdot {\bf v}\right) + \left(q - 1\right)(\nabla V_{ext}) \nabla \cdot {\bf v} \\ + \nabla \left({\bf v} \cdot \nabla V_{ext}\right)
\end{multline}
in terms of the adiabatic compressibility and the polytropic coefficient $q$. 
Equation \eqref{Eq:eqvfinal} coincides with the result derived in \cite{Griffin1997} for a 3D ideal Bose gas for $T \ge T_c$ with $q=5/3$. At  $T = 0$, where $\gamma=1$, it  reduces to
\begin{equation}
\label{Eq:zero_T}
m\omega^2{\bf v} = - (q - 1)\nabla\left[\mu(n)\nabla \cdot {\bf v}\right] + \nabla\left[{\bf v} \cdot \nabla V_{ext}\right]
\end{equation}
where we have used the $T=0$ expression \eqref{Eq:mu} for the chemical potential $\mu(n)$.

In the form  (\ref{Eq:eqvfinal})  (and (\ref{Eq:zero_T}) at $T=0$) the equation for the velocity field admits a useful class of analytical solutions that will be discussed in the following Sections where
 we will consider an  external potential of the shape
\begin{equation}
\label{Eq:anisotropic_potential2D}
V_{\mathrm{ext}}(x,y) = \frac{1}{2}m \left(\omega_x^2 x^2 + \omega_y^2 y^2  \right)
\end{equation}
for the investigation of the collective oscillations of 2D configurations and of the form 
\begin{equation}
\label{Eq:anisotropic_potential1D}
V_{\mathrm{ext}}(z) = \frac{1}{2}m \omega_z^2 z^2 
\end{equation}
in the case of 1D configurations.

\section{Surface modes}
\label{Sec:surface}

For divergency free oscillations ($\nabla \cdot {\bf v} = 0$), also called surface modes,
the hydrodynamic equation \eqref{Eq:eqvideal} takes the  simplified form

\begin{equation}
m\omega^2 {\bf v}=  \nabla \left({\bf v} \cdot \nabla V_{ext}\right) 
\label{Eq:HDsurface}
\end{equation}
independent of the equation of state, quantum statistics and temperature. 

In the case of 2D isotropic trapping ($\omega_x = \omega_y \equiv \omega_\perp$)  an important class of solutions of Eq. \eqref{Eq:HDsurface} is given by the irrotational choice \cite{Stringari1996}:
\begin{equation}
\label{Eq:surface_ansatz}
{\bf v}(r_{\perp}, \phi) \propto \nabla \left[r_\perp^{|m|} e^{\pm i m \phi}\right],
\end{equation}
where $m$ is $z$-th component of angular momentum.
The resulting eigenfrequencies take the form
\begin{equation}
\label{Eq:surface}
\omega^2(m) = \omega^2_\perp |m| \; .
\end{equation}

Important examples of divergency free oscillations, systematically investigated in harmonically trapped atomic gases, are the dipole ($\omega_D=\omega_\perp$) and the quadrupole ($\omega_Q=\sqrt{2}\omega_\perp$) oscillations. The former case is model independent, due to Kohn's theorem. The latter case is particularly interesting because the predicted frequency differs from the collisionless ideal gas value  $2\omega_\perp$ and its investigation can then be used as a direct test of the achievement of the hydrodynamic nature of the regime considered.

In the case of anisotropic potentials ($\omega_x\ne \omega_y$) an important example of divergency free  oscillations is described by the  ansatz: 
\begin{equation}
\label{Eq:scissors}
{\bf v}(x, y) \propto \nabla(x y)
\end{equation}
yielding the result
\begin{equation}
\label{Eq:scissors2}
\omega^2_S = \omega_x^2 + \omega_y^2
\end{equation}
for the collective frequency.
This is the  so-called scissors mode corresponding to an  oscillating rotation of the atomic cloud in the $x$-$y$ plane. It was first predicted by Gu\'ery-Odelin and Stringari (1999) \cite{Guery-Odelin1999} and experimentally observed in both 3D Bose \cite{Marago2000} and Fermi superfluids \cite{Wright2007}. This mode preserves the shape of the gas during the oscillation and for this reason is independent of the equation of state. It has the same value in 3D as well as in 2D configurations described by the hydrodynamic formalism.

\section{Compressional oscillations  in pancake and 2D isotropic traps}
\label{Sec:2D_isotropic}

Let us first consider the superfluid zero temperature regime for which, as already pointed out in the Section \ref{Sec:hydrodynamics}, the irrotationality condition for the velocity field holds. For systems whose equation of state is characterized by the polytropic density dependence \eqref{Eq:mu}, the velocity field  obeys the hydrodynamic Eq. \eqref{Eq:zero_T}. For isotropic 2D traps one can easily check that this equation is solved  by the following irrotational ansatz for the velocity field \cite{Stringari1998}:
\begin{equation}
\label{Eq:isotropic_ansatz}
{\bf v}(r_{\perp}, \phi) \propto \nabla\left[(r_\perp^{2n}+....)r_\perp^{|m|} e^{\pm i m \phi}\right]
\end{equation}
where $n$ fixes the number of the radial nodes of the density modulations occurring during the oscillation and $m$ is $z$-th component of angular momentum. The resulting hydrodynamic eigenfrequencies take the form
\begin{equation}
\label{Eq:omega_isotropo}
\omega^2(n, m, q) = \omega_\perp^2[2n + |m| + 2n(q - 1)(n + |m|)].
\end{equation}
Result \eqref{Eq:omega_isotropo} can be applied to a rich variety of  physical situations whose equation of state is simply incapsulated  in the polytropic $q$ index and it differs from the predicted values of the ideal gas model in the collisionless regime
\begin{equation}
\label{Eq:omega_ideal}
\omega_{cl}(n, m) = \omega_\perp(2n + |m|).
\end{equation} 
Moreover, result \eqref{Eq:omega_isotropo} should be compared with the hydrodynamic dispersion law holding in the presence of a 3D isotropic  harmonic potential ($\omega_x=\omega_y=\omega_z \equiv \omega_{ho}$)
\begin{equation}
\label{Eq:omega_isotropo3D}
\omega^2(n, l, q) = \omega_{ho}^2[2n + l + (q - 1)(2n^2 + 2nl + n)]
\end{equation}
where $l$ is the angular momentum. Eq. \eqref{Eq:omega_isotropo3D} reduces to the results of \cite{Stringari1996} and \cite{Baranov2000, Bruun1999, Minguzzi2001} in the case of the weakly interacting Bose ($q = 2$) and Fermi ($q = 5/3$) gases, respectively. Result \eqref{Eq:omega_isotropo}
reduces to the dispersion \eqref{Eq:surface} of surface modes in the case $n = 0$. 

The most important solution accounted for by \eqref{Eq:omega_isotropo} is the 2D breathing or monopole mode ($n = 1$ and $m = 0$), for which one finds the result
\begin{equation}
\label{Eq:monopole} 
\omega^2_M(q) = 2q\omega^2_\perp .
\end{equation}
Differently from the surface modes discussed in the previous Section, the frequency of the breathing  mode depends explicitly on the polytropic index $q$ and is consequently sensitive to the equation of state obeyed by the gas. The frequency \eqref{Eq:monopole} of the 2D breathing mode differs from the one in the 3D isotropic case (see Eq.\eqref{Eq:omega_isotropo3D}, $n = 1$ and $l = 0$) which yields  the value $\omega^2_M(q) = (3q-1) \omega^2_{ho}$. 
 
For the unitary Fermi gas  in the pancake regime ($q = 3/2$), one gets the value  $\omega_M = \sqrt{3}\omega_\perp$. Actually this value turns out to be  independent of temperature, as a consequence  of the fact that, at unitarity, the scaling solution for the breathing mode is exactly satisfied by the hydrodynamic equations at all temperatures \cite{Hou2013}.  Notice that this result differs from the value $\omega_M = 2\omega_{ho}$  holding  for the unitary Fermi gas with 3D isotropic trapping ($q = 5/3$). In this latter case  an exact solution of the time dependent Schr\"odinger equation is available for the monopole breathing mode, independent not only of temperature but also of  the amplitude of the oscillation and holding in all collisional regimes (scale invariance) \cite{Castin2004}. A similar situation holds also for the  2D regime of a weakly interacting Bose gas  ($q = 2$) in the presence of isotropic trapping where one also finds the result $\omega_M= 2\omega_\perp$, independent of temperature and of the amplitude of oscillations \cite{Pitaevskii1996, Pitaevskii1997}. In the case of the 2D Fermi gas the result $\omega_M = 2\omega_\perp$, following from the $q = 2$ value of the polytropic coefficient, holds both in the BEC and BCS regime and, with high accuracy, along the whole crossover, revealing an apparent scale invariance \cite{Taylor2012, Levinsen2015}, as proven experimentally in \cite{Vogt2012}.

The temperature dependence of the frequency of the breathing mode takes instead place for the pancake weakly interacting Bose gas. At $T=0$ ($q = 5/3$) one finds $\omega_M = \sqrt{10/3}\omega_\perp$ \cite{Stringari1996}. This result differs from the value obtained at high temperature, where the thermodynamic behaviour of the gas can be approximated by the ideal Bose gas ($q=3/2$) and the hydrodynamic frequency takes the value $\omega_M = \sqrt{3}\omega_{\perp}$.
The  frequency of the  monopole mode of  the pancake Bose gas is then expected to exhibit a temperature dependence. 
The temperature dependence of this mode was pointed out in the first experiments on the collective oscillations carried out at Jila \cite{Jin1997} where, however, due to the small number of atoms, the  system enters soon the collisionless regime for $T > T_c$, rather than the collisional one.  Achieving the hydrodynamic condition $\omega \tau \ll 1$ is in general  a difficult task in weakly interacting Bose gases, due the small available values of the density of the gas.  A simple estimate, based on the classical evaluation of the collisional time, yields, for frequencies of the order of the  trapping frequency $\omega_{ho}$, the condition \cite{Dalfovo1999} $l_{mph} \ll  R_T$, where $l_{mph}=(n\sigma)^{-1 }$ is the mean free path. Here   $n$ is the local 3D density,  $\sigma=8\pi a^2$ is the s-wave cross section, while $R_T= \sqrt{2k_BT/m\omega^2_{ho}}$ is the   thermal radius.  

A summary of the frequencies of the  breathing mode in the most relevant cases discussed in this section is presented in  Tables \ref{Tab:monopoleTF} and \ref{Tab:monopoledeep}.

\begin{table}
\caption{Hydrodynamic frequencies of the breathing mode in  different  LDA regimes. The values reported for the Unitary Fermi gas hold also for the ideal Bose gas above $T_c$.}
\label{Tab:monopoleTF}
\begin{tabular}{ccc}
 \hline\noalign{\smallskip}
 \hline\noalign{\smallskip}
&\textbf{Bose gas (T = 0)} & \textbf{Unitary Fermi gas} \\
\noalign{\smallskip}\hline\noalign{\smallskip}
\textbf{3D isotropic} & $\sqrt{5}\omega_{ho}$ & $2\omega_{ho}$    \\
\noalign{\smallskip}\hline\noalign{\smallskip}
\textbf{pancake (LDA)} &   $\sqrt{10/3}\omega_\perp$  &   $\sqrt{3}\omega_\perp$   \\
\noalign{\smallskip}\hline\noalign{\smallskip}
\textbf{cigar (LDA)} &   $\sqrt{5/2}\omega_z$  &   $\sqrt{12/5}\omega_z$   \\
\noalign{\smallskip}\hline 
\noalign{\smallskip}\hline 
\end{tabular}
 \end{table}

\begin{table}
\caption{Hydrodynamic frequencies of the breathing mode for a Bose gas in low dimensions and tight trapping regimes. The 2D mean field value $\omega_M = 2\omega_\perp$ holds also for the interacting Fermi gas in the BCS and BEC limits.}
\label{Tab:monopoledeep}
\begin{tabular}{ccc}
 \hline\noalign{\smallskip}
 \hline\noalign{\smallskip}
&\textbf{T = 0} & \textbf{high T} \\
\noalign{\smallskip}\hline\noalign{\smallskip}
\textbf{2D mean field} & $2\omega_\perp$ & $2\omega_\perp$    \\
\noalign{\smallskip}\hline\noalign{\smallskip}
\textbf{1D mean field} &   $\sqrt{3}\omega_z$  &   $2\omega_z$   \\
\noalign{\smallskip}\hline\noalign{\smallskip}
\textbf{1D Tonks-Girardeau} &   $2\omega_z$  &   $2\omega_z$   \\
\noalign{\smallskip}\hline 
\noalign{\smallskip}\hline 
\end{tabular}
 \end{table}

\section{Coupled compressional and surface modes in anisotropic traps}
\label{Sec:coupled}

In the presence of anisotropic trapping ($\omega_x\ne \omega_y$),  the breathing and the quadrupole oscillations discussed in the previous Sections are coupled. The coupling is  accounted for by the ansatz
\begin{equation}
\label{Eq:qm}
{\bf v}(x, y) =\nabla(\alpha x^2 + \beta y^2),
\end{equation}
where the relative value of the parameters $\alpha$ and $\beta$ have to be determined by solving the hydrodynamic equation \eqref{Eq:eqvfinal}.  Since  $\nabla\left(\nabla \cdot {\bf v}\right) = 0$ the frequency of the coupled modes, for a given value of the polytropic coefficient  $q$, are given by
\begin{multline}
\label{Eq:coupled2}
\omega_{\pm}^2(q) = \frac{(q + 1)}{2}(\omega_x^2 + \omega_y^2)  \\
\pm \frac{\sqrt{(q + 1)^2(\omega_x^2 + \omega_y^2)^2 - 16q\omega_x^2 \omega_y^2}}{2} \; .
\end{multline}

In the isotropic limit ($\omega_x=\omega_y = \omega_\perp$)  Eq. \eqref{Eq:coupled2} reproduces the results (\ref{Eq:monopole}) and (\ref{Eq:surface}) for the monopole and the quadrupole, respectively, derived in the previous Sections.  In the case of  the 2D mean field regime, where the polytropic coefficient takes the value $q = 2$, result (\ref{Eq:coupled2}) was derived in \cite{Ghosh2000} and in \cite{Baur2013} for bosons and fermions, respectively.

Result (\ref{Eq:coupled2}) gives general predictions for the collective frequencies in 2D configurations  for arbitrary values of the deformation of the trap. It is interesting to consider the limit of highly deformed 2D trapping potentials. For  $\omega_x \ll \omega_y$  the lower solution takes the form 
\begin{equation}
\label{Eq:omega-}
\omega^2_- = \frac{4q}{(q + 1)}\omega_x^2,
\end{equation}
while  the upper solution is given by
\begin{equation}
\label{Eq:omega+}
\omega^2_+ = (q + 1) \omega^2_y
\end{equation}
Of course a symmetric result takes place in the opposite limit $\omega_x \gg \omega_y$.

In the 2D mean field case ($q=2$) \eqref{Eq:omega-} provides the result $\omega_-=\sqrt{8/3}\omega_x$, in agreement with the findings of \cite{Ghosh2000}. In the pancake regime \eqref{Eq:omega-} instead reduces to the frequency of the axial breathing mode of a cigar configuration (see next Section). 


Equation \eqref{Eq:omega+} instead coincides with the  solution of the 3D hydrodynamic equations for triaxial harmonic trapping \cite{Pitaevskii2003} in the intermediate regime $\omega_z\gg \omega_y\gg \omega_x$. 

\section{Collective frequencies in cigar and 1D traps}
\label{Sec:1D}

In this Section we complete the discussion on the collective frequencies of harmonically trapped gases considering one dimensional configurations. 

Similarly to the 2D case, two different regimes can be considered also in 1D. The first one, called cigar regime, corresponds to systems described locally by the 3D equation of state, but the radial confinement is enough tight to ensure a 1D nature to  the low energy dynamic behavior. This configuration is particularly suited to investigate the propagation of sound \cite{Andrews1997, Joseph2007, Horikoshi2010}. 

A second case is the deep 1D (hereafter simply called 1D) regime where the radial motion is frozen (for a general review on 1D systems see, for example, \cite{Giamarchi2004, Giamarchi2014}). The role of correlations is particularly important in this regime and, in the case of Bose gases at $T=0$, is described by Lieb-Lininger theory \cite{Lieb1963}. 
A key question in 1D is the role of thermalization. From this point of view the comparison between experiments and the predictions of hydrodynamic theory (which assumes local thermalization at finite temperature) would be particularly insightful.  Another important feature of 1D systems is the absence of phase transitions at finite temperature \cite{Landau1980}.  

The equations of hydrodynamics derived in Sect. \ref{Sec:hydrodynamics} become particularly simple in 1D, the velocity field being a function of the variable $z$ and all the gradients acting only on the $z$-direction.
Looking for solutions of the form
\begin{equation}
v= z^k + \alpha_{k-2}z^{k-2} + ...
\label{v1D}
\end{equation}
one finds that the hydrodynamic equation \eqref{Eq:eqvfinal}, in the presence of the polytropic equation of state \eqref{Eq:q}, admits simple analytic solutions in the presence of the harmonic trapping \eqref{Eq:anisotropic_potential1D}. 

At $T=0$, where one can conveniently use Eq. \eqref{Eq:zero_T} with the chemical potential given by (\ref{LDAplanar}), the dispersion law takes the form 
\begin{equation}
\omega^2(k,q)= (k+1)\left[(q-1)\frac{k}{2}+1\right]\omega_z^2
\label{omega1DT=0} \; .
\end{equation}
Equation (\ref{omega1DT=0}) was first derived in \cite{Menotti2002} in the context of 1D Bose gases. 

In the classical limit of high temperature, where one can use the equation of state $P_{1D}=n_{1D}k_BT$, the hydrodynamic equation \eqref{Eq:eqvfinal} gives rise instead to the following dispersion relation
\begin{equation}
\omega^2(k,q)= \left(qk+1\right)\omega_z^2
\label{omega1DTcl}
\end{equation}

Result (\ref{omega1DT=0}) should be also compared with the dispersion
\begin{equation}
\omega_{cl}(k) = (k+1) \omega_z
\label{cl1D}
\end{equation}
holding in the collisionless regime of a non interacting  1D gas. 

Let us first consider the case of the cigar unitary Fermi gas. In this case the polytropic coefficient is equal to $q = 7/5$ at all temperatures and the $k=1$ hydrodynamic  solution, corresponding to the lowest breathing (LB) mode, is equal to $\omega_{LB} = \sqrt{12/5}\omega_z$ \cite{Heiselberg2004,Stringari2004} at both zero and finite temperature. 
The situation is different for the higher nodal modes ($k > 1$) where the hydrodynamic frequencies at $T=0$ and in the classical limit are different revealing an interesting temperature dependence \cite{Hou2013_2} that was investigated experimentally in \cite{Tey2013}  in good agreement with the predictions of theory. The case of the cigar Bose gas is different. In fact in this case the polytropic coefficient depends on temperature, being equal to $q = 3/2$ at $T=0$ and to $q = 7/5$ in the classical limit. As a result, the frequency of the $k=1$ lowest breathing mode takes the value $\omega_{LB} = \sqrt{5/2} \omega_z$ \cite{Stringari1996,Stringari1998} while in the high temperature classical limit one finds the smaller hydrodynamic value $\omega_{LB} = \sqrt{12/5}\omega_z$. In the collisionless limit one instead finds the value $2\omega_z$. The $T = 0$ value $\sqrt{5/2} \omega_z$ was  measured experimentally in \cite{Stamper-Kurn1998} in excellent  agreement with the prediction of theory. At finite temperature these authors found that the frequency  slightly drops below the low temperature limit. At even higher temperatures they observed a significant  increase of $\omega$, likely due to the breakdown of the hydrodynamic condition $\omega \tau \ll 1$.

The comparison between the lowest compressional mode in the cigar geometry ($k=1$) and the lowest solution (\ref{Eq:omega-}) holding in the pancake geometry with highly anisotropic 2D trapping, allows for the non trivial relationship
\begin{equation}
q_{cigar}+1= \frac{4 q_{pancake}}{q_{pancake}+1}
\label{qcp}
\end{equation}
between the polytropic coefficients in the cigar and pancake geometries. The relationship is confirmed by the results reported in Table \ref{Tab:qTF}.

The behavior of the collective frequencies in the deep 1D regime is also very interesting. In the case of bosons, one should distinguish between the mean field case, where $q=2$ at $T=0$, and the Tonks-Girardeau (TG) limit \cite{Girardeau1960}, where $q=3$. In the former case the frequency of the $k=1$ breathing mode takes the value $\omega_{LB} = \sqrt{3}\omega_z$, while in the TG limit one finds $\omega_{LB} = 2\omega_z$. The behaviour of the lowest breathing frequencies, in the intermediate regimes between the mean-field and the TG limits, was theoretically investigated in \cite{Menotti2002} and experimentally observed in \cite{Haller2009}. At high temperatures, both regimes predict the frequency $\omega_{LB} = 2\omega_z$ for the same $k = 1$ mode.
 One then concludes that, differently from the 1D mean field, the frequency of the breathing mode in the TG regime is temperature independent. The result is not surprising since in the TG limit bosons behave like 1D non interacting fermions where scaling invariance applies. It is worth noting, however, that the hydrodynamic  frequencies of the higher nodal modes depend on temperature even in the TG limit. For example, the $k=2$ mode has frequency  $\omega_{k=2} = 3\omega_z$ at $T=0$ and $\omega_{k=2} = \sqrt{7}\omega_z$ in the classical limit. The temperature dependence of the breathing mode in the 1D mean field has been recently measured in \cite{Fang2014}. 

\section{Conclusions}
\label{Sec:Conclusion}
In this paper we have derived a unified description of the discretized collective oscillations of quantum gases in different conditions of trapping and dimensionality. 

 A major result of this work is given by the derivation of a general  hydrodynamic equation for the velocity field (Eq. \eqref{Eq:eqvideal}) depending solely on the adiabatic and isothermal compressibilities of the gas. This equation takes a particularly simple form in the case of systems exhibiting a polytropic equation of state, characterized by  a power law dependence of the pressure on the density ($P\propto n^q$),  for a fixed value of the entropy.  The polytropic equation of state characterizes a significant class of many body configurations  of either bosonic and fermionic nature. 

Analytic  results for the collective frequencies have been calculated for pancake and cigar configurations, where the equation  of state can be obtained from the 3D equation of state using the local density approximation, as well as in the deep 2D and 1D regimes where the motion is instead frozen along the direction of the confinement.  Special emphasis is given to the lowest breathing mode, whose frequency has been shown to depend explicitly on the value of the polytropic coefficient  $q$, and to its coupling with the quadrupole oscillations in the presence  of anisotropic external potentials.  

We have emphasized, in particular, the comparison between the $T=0$ results, characterizing the behavior of superfluids, and the high temperature behavior, both in the hydrodynamic and collisionless regime. 
In this respect an interesting perspective of the present work is given by the study of the collective frequencies in 1D configurations at finite temperature which could provide  a useful test of the equation of state  of strongly interacting gases \cite{Yang1969, Yang1970, Kheruntsyan2005} as well as the achievement of the collisional (hydrodynamic) condition at finite temperature.

Our results can be also used as a starting point to calculate the collective frequencies in systems whose  equation of state  deviates from  the polytropic law and the solution of the hydrodynamic equations can be handled using a perturbative approach. This procedure, first employed to estimate the frequency shifts of dilute Bose gases caused by beyond mean field effects \cite{Pitaevskii1998}, was developed in a systematic way by Astrakharchik \cite{Astrakharchik2005} in a variety of configurations of different dimensionality and quantum statistics. It was recently  applied by Merloti et al. \cite{Merloti2013} to explore the breakdown of scale invariance in a quasi-two-dimensional Bose gas due to the presence of the third dimension.  

\appendix
\section{Equation of state and polytropic coefficient}
\label{Sec:EOS}

In this Appendix we identify the polytropic coefficient $q$ characterizing the equation of state \eqref{Eq:q}  of a uniform gas in different low dimensional configurations   and for different quantum statistics (see Tables \ref{Tab:qTF} and \ref{Tab:qdeep}). 

Let us  first consider  the ideal situation of zero temperature.
As already discussed in Section \ref{Sec:polytropic equation}, a first interesting two dimensional regime is the  so called pancake  where the system keeps, locally, its 3D nature in the sense that the chemical potential is much larger than axial oscillator energy $\mu \gg \hbar \omega_z$. 
In this case we can  apply the local density approximation (LDA) along the axial direction and write the chemical potential in the form
\begin{equation}
\mu_0=\mu(n)+\frac{m}{2}\omega^2_zz^2 \; 
\label{LDAplanar}
\end{equation}
which allows us to determine the $z$-dependence of the density profile. The value of $\mu_0$  coincides with  the chemical potential $\mu(n)$ calculated at $z=0$ and   is fixed by the normalization condition $n_{2D}=\int dz n({\bf r})$. The quantity $\mu_0$ plays consequently the role of the 2D chemical potential and exhibits an  explicit dependence on the 2D density  $n_{2D}$. This dependence characterizes the   2D equation of state and, for simplicity, we will omit the suffix $0$ in $\mu_0$ and use the simpler notation $\mu(n_{2D})$ to characterize the equation of state of the two-dimensional gas.

The 2D equation of state of a pancake gas is easily derived in the case of dilute Bose gas where the 3D equation of state has the form $\mu=gn$  and the use of the LDA procedure (\ref{LDAplanar}) gives the result \cite{Pitaevskii2003}
\begin{equation}
\mu^B(n_{2D}) = \left( \frac{3\pi \hbar^2 \omega_z a n_{2D}}{\sqrt{2m}}\right)^{2/3} \; ,
\label{pancakeB}
\end{equation}
characterized by  the value $q=5/3$ for the polytropic coefficient.

An analogous calculation can be  carried out for a pancake Fermi gas at unitarity, where the 3D equation of state has the form $\mu = \xi_B \frac{\hbar^2}{2m}(3\pi^2 n)^{2/3}$, with $\xi_B$ the Bertsch parameter, yielding  \cite{Pitaevskii2003}
\begin{equation}
\mu^F(n_{2D}) = \left(2 \pi \xi_B^{\frac{3}{2}}\omega_z \frac{\hbar^3}{m}n_{2D}\right)^{1/2} \; ,
\label{pancakeF}
\end{equation}
and hence  the  value $q=3/2$ for the polytropic coefficient. 

Similar results can be obtained in cigar 1D-like configurations where the gas is harmonically trapped along the 
$x-y$ directions. In this case,
 after integrating the radial   3D Thomas-Fermi profile, one finds  \cite{Pitaevskii2003}
\begin{equation}
\mu^B(n_{1D})=2\hbar \omega_\perp (an_{1D})^{1/2}
\label{muB1}
\end{equation}
and 
\begin{equation}
\mu^F(n_{1D}) = \left(\frac{\xi_B\hbar^2}{2m}\right)^{3/5}\left(\frac{15}{4}\pi m \omega_\perp^2n_{1D}\right)^{2/5}
\label{muF1}
\end{equation}
 for the chemical potential of the cigar Bose and unitary Fermi gas respectively where $
n_{1D}=\int dxdy n({\bf r})$. 
From the above equations one derives the values $q= 3/2$ and $q=7/5$ for the polytropic coefficients in the Bose and unitary Fermi cigars, respectively which are reported in the Table \ref{Tab:qTF}.

The results $q=3/2$ and $q = 7/5$ for the polytropic coefficients of the pancake and cigar Fermi gas at unitarity holds also at finite temperature. In fact the 3D result for the equation of state, given by Eq. (\ref{Eq:q}) with $q= 5/3$, can be usefully rewritten  in the form  \cite{Ho2004,Hou2013_2}
\begin{equation}
P_{3D}(x, T) = f_p(x) \frac{k_BT}{\lambda_T^3}
\label{P3DT}
\end{equation}
and 
\begin{equation}
n_{3D}(x, T) = \frac{f_p'(x)}{\lambda_T^3} = \frac{f_n(x)}{\lambda_T^3} 
\label{n3DT}
\end{equation}
where the temperature dependence of the 3D pressure and density follows from dimensionality arguments. We have here introduced the thermal wavelength $\lambda_T = \sqrt{2\pi\hbar^2/mk_BT}$, the dimensionless functions  $f_p(x)$ and its  derivative $f_n(x)= f_p'(x)$. The dimensionless parameter $x=\mu/k_BT$, fixed by the ratio between the chemical potential and the temperature of the  gas, determines  the entropy per particle $\bar{s}$ according to the  relationship 
\begin{equation}
\frac{\bar{s}(x)}{k_B} = \frac{5}{2}\frac{f_p(x)}{f_n(x)} - x 
\label{sbar}
\end{equation}
The above results reflect the universality of the 3D uniform Fermi gas at unitarity.

By applying  the local density approximation (\ref{LDAplanar}) along the $z$-direction in the pancake geometry and along  the $x-y$ plane in the cigar geometry, after integration of the density and of the pressure profile along the same directions, from Eq. (\ref{P3DT}) and Eq. (\ref{n3DT}) one easily finds a new temperature dependence of the 2D and 1D pressure and density for a given value of the entropy, fixed by the ratio $x=\mu/k_BT$. For the pancake configuration one finds
\begin{equation}
P_{2D}(x_{2D}, T) = \frac{k_BT}{\lambda_T^3}\sqrt{\frac{2k_BT}{m\omega_z^2}}F^{2D}_p(x_{2D})
\label{P2DT}
\end{equation}
and 
\begin{equation}
n_{2D}(x_{2D}, T) = \frac{1}{\lambda_T^3}\sqrt{\frac{2k_BT}{m\omega_z^2}}F^{2D}_n(x_{2D})
\label{n2DT}
\end{equation}
while for the cigar 1D configuration the result is \cite{Hou2013_2}
\begin{equation}
P_{1D}(x_{1D}, T) = \frac{k_BT}{\lambda_T^3}\frac{2k_BT}{m\omega_\perp^2}F^{1D}_p(x_{1D})
\label{P1DT}
\end{equation}
and 
\begin{equation}
n_{1D}(x_{1D}, T) = \frac{1}{\lambda_T^3}\frac{2k_BT}{m\omega_\perp^2}F_n^{1D}(x_{1D})
\label{n1DT}
\end{equation}
holding at all temperatures. From the above results one easily finds, using Eq. (\ref{Eq:q})  the values $q=3/2$ and $q=7/5$ for the polytropic coefficient in the 2D and 1D cases, independent of temperature. In Eq. (\ref{P2DT}) and (\ref{n2DT}), we have defined the integrated functions $F^{2D}_{n, p}(x_{2D}) = \int_{- \infty}^{+ \infty}dz' f_{n, p}(x')$ with $x_{2D} = \mu_0/k_BT$, while in Eq. (\ref{P1DT}) and (\ref{n1DT}) the analogous 1D integrated functions are $F^{1D}_{n, p}(x_{1D}) = 2\pi\int_{0}^{+ \infty}dr_\perp' r_\perp' f_{n,p}(x')$ with $x_{1D} = \mu_0/k_BT$. The same results for the values of $q$ hold also for the ideal Fermi gas as well as for the ideal Bose gas above $T_c$ where the thermodynamic functions can be written in the same form as for the unitary Fermi gas, the functions $f_n$ and $f_p$ being of course different. 

So far we have considered the pancake and cigar geometries where the local density approximation allows us to safely use the 3D equation of state locally. 
The situation is different  in the opposite regime of tight axial or radial confinement where the motion is frozen to the lowest harmonic oscillator wave function along the tight directions.  At zero temperature the condition for being in these  deep 2D and 1D regimes is given by
$\mu \ll \hbar \omega_z$ and $\mu \ll \hbar \omega_\perp$, respectively. At high temperature the conditions are, instead, $k_BT \ll \hbar \omega_z$ and $k_BT \ll \hbar \omega_\perp$, respectively. 
At $T=0$ these low dimensional regimes are    easily described  in the case of a weakly interacting Bose gas where the use of Gross-Pitaevskii theory yields the simple results for the 2D and 1D equations of state
\begin{equation}
\mu^B(n_{2D}) =  g_{2D} n_{2D} 
\label{2DGP}
\end{equation}
and
\begin{equation}
\mu^B(n_{1D}) =  g_{1D} n_{1D} 
\label{1DGP}
\end{equation}
apart from unimportant additional constant terms. In Eq. (\ref{2DGP}) the 2D coupling constant is given by  \cite{Pitaevskii2003}
\begin{equation}
g_{2D}= \sqrt{8\pi} \frac{\hbar^2}{m} \frac{a}{a_z}
\label{g2D}
\end{equation}
and is fixed by the ratio $a/a_z$  between the 3D s-wave scattering length and the axial harmonic oscillator length $a_z=\sqrt{\hbar/m\omega_z}$, while in the 1D case one finds the result  \cite{Pitaevskii2003}
\begin{equation}
g_{1D}= 2a \hbar \omega_\perp
\label{g1D}
\end{equation}
Result (\ref{2DGP}) for the equation of state of the 2D Bose gas holds only in the limit of weak coupling ($a\ll a_z$), ensuring the absence of quantum anomaly effects \cite{Papoular2015}. 
The 1D equation of state (\ref{1DGP}) instead holds provided $n_{1D}a^2_\perp/a \gg 1$. 
In both cases (also called 2D and 1D mean field regimes) one identifies the value $q=2$ for the  polytropic coefficient.   Both  the 2D and 1D  mean field regimes have been achieved experimentally in \cite{Desbuquois2012} and   \cite{Kinoshita2004, Paredes2004}, respectively. 

In the 1D case, extensive experimental and theoretical work has been done also to explore regimes beyond the mean field condition $n_{1D}a^2_\perp/a \gg 1$ where the many-body properties of the gas are described by  Lieb-Lininger theory.   In the limit $n_{1D}a^2_\perp/a \ll 1$, the gas enters the limit of impenetrable bosons also called Tonks-Girardeau regime where the gas acquires a Fermi like behavior and its equation of state exhibits a  quadratic density dependence \cite{Girardeau1960}:
\begin{equation}
\mu^{TG}(n_{1D}) =  \pi^2 \frac{\hbar^2}{2m}n^2_1 
\label{1DTG}
\end{equation}
In this regime the value of the polytropic coefficient is $q=3$.

In the case of interacting Fermi gases, the theoretical description  of the lower dimensional regimes is more difficult with respect to the Bose case. In two dimensions  one can still identify a BCS-BEC crossover like in the 3D case. However, in the two-body problem  the presence of a resonance always corresponds to the occurrence of a bound state, differently from what happens in 3D where this is ensured only for positive values of the 3D scattering length. 
Far from the resonance the chemical potential of the 2D Fermi gas is described by a linear dependence on the density, both on the BEC side, where the gas is described by a 2D system of bosonic molecules, and on the deep BCS limit, where the gas approaches the ideal Fermi gas behaviour.  In both cases the value of the polytropic coefficient is $q=2$.  The same value of $q$ characterizes the equation of state of the classical gas at high temperature. 

Tables \ref{Tab:qTF} and \ref{Tab:qdeep} summarize the main results of this Section, reporting the values of the polytropic coefficients $q$ in a class of interesting configurations. 

\section{Hydrodynamic equations in the presence of an external potential}
\label{Sec:appendix_wave}

In this Appendix we will derive the most relevant hydrodynamic equations used in the paper to calculate the collective frequencies in the presence of external harmonic trapping. 

In order to derive Eq. \eqref{Eq:eqvideal}, we first take the time derivative of Eq. \eqref{Eq:j} yielding, in the limit of small amplitude oscillations, 
\begin{equation}
\label{Eq:j_t}
mn_0\frac{\partial^2}{\partial t^2}{\bf v} = - \nabla \frac{\partial }{\partial t}P - \nabla V_{ext}\left(\frac{\partial }{\partial t}n\right) \; .
\end{equation}
The time derivative of the pressure can be written as:
\begin{equation}
\label{Eq:P_t}
\frac{\partial P}{\partial t} = \left(\frac{\partial P}{\partial n}\right)_{T}\frac{\partial n}{\partial t} + \left(\frac{\partial P}{\partial T}\right)_{n}\frac{\partial T}{\partial t}
\end{equation}
and requires the knowledge of the time derivative of the temperature. 

In order to calculate $\partial T/\partial t$ it is convenient 
to rewrite the equation for the entropy density \eqref{Eq:continuity_entropy} in terms of the entropy per particle $\bar{s} = s/n$. By using the equation of continuity \eqref{Eq:continuity2}, it is immediate to find the equation 
\begin{equation}
\label{Eq:continuity_entropy_final}
\frac{\partial \bar{s}(\bold{r}, t)}{\partial t} = -\bold{v}(\bold{r}, t) \nabla \bar{s}_0(\bold{r}) =  \frac{1}{n_0^2} \left(\frac{\partial P}{\partial T}\right)_n {\bf v} \cdot \nabla n_0
\end{equation}
where, in deriving the second equality, we have used 
the thermodynamical relation 
\begin{equation}
Td\bar{s} = c_vdT - \frac{T}{n^2}\left(\frac{\partial P}{\partial T}\right)_ndn
\label{Maxwell}
\end{equation}
applied to equilibrium ($dT = 0$),
where 
\begin{equation}
\label{Eq:c_v}
c_v = \frac{T}{n_0^2}\left(\frac{\partial P}{\partial T}\right)_{n} \left(\frac{\partial n}{\partial T}\right)_{\bar{s}} 
\end{equation}
is the specific heat at constant volume.

On the other hand, by considering $\bar{s}$ as a function of density and temperature, one can also write
\begin{equation}
\label{Eq:s_t_2}
\frac{\partial \bar{s}}{\partial t} = \left(\frac{\partial \bar{s}}{\partial n}\right)_T\frac{\partial n}{\partial t} + \left(\frac{\partial \bar{s}}{\partial T}\right)_n\frac{\partial T}{\partial t} \; .
\end{equation}
Using the equation of continuity \eqref{Eq:continuity2} and  thermodynamic relation \eqref{Eq:continuity_entropy_final}, one finally obtains the useful equation 
\begin{equation}
\label{Eq:T_t2}
\frac{\partial T({\bf r}, t)}{\partial t} = - \frac{T}{c_v}\left(\frac{\partial P}{\partial T}\right)_n \frac{\nabla \cdot {\bf v({\bf r}, t)}}{n_0}
\end{equation}
for the time derivative of the temperature, holding also in the presence of external trapping.

Using the equilibrium condition in the Eq. \eqref{Eq:j}
\begin{equation}
\label{Eq:nabla_P}
\nabla P_0 = \left(\frac{\partial P_0}{\partial n_0}\right)_T \nabla n_0 = - n_0 \nabla V_{ext}
\end{equation}
and the thermodynamic relation
\begin{equation}
\label{Eq:P_n}
\left(\frac{\partial P}{\partial n}\right)_{\bar{s}} = \left(\frac{\partial P}{\partial n}\right)_T + \left(\frac{\partial P}{\partial T}\right)_n \left(\frac{\partial T}{\partial n}\right)_{\bar{s}}
\end{equation}
relating the adiabatic and isothermal compressibilities,
it is  easy to recast the hydrodynamic equation (\ref{Eq:j_t}) in the useful form 
\begin{multline}
m\omega^2 {\bf v}= -\nabla \left[\left(\frac{\partial P}{\partial n}\right)_{\bar{s}}(\nabla \cdot {\bf v})\right] + (\gamma - 1) (\nabla V_{ext})(\nabla \cdot {\bf v})  \\
+ \nabla \left({\bf v} \cdot \nabla V_{ext}\right) 
\label{Eq:eqvideal_appendix}
\end{multline}
where $\gamma$ is the adiabatic coefficient defined by (\ref{Eq:gamma}) and 
we have considered velocity fields oscillating as
${\bf v}({\bf r},t)={\bf v}({\bf r})e^{-i\omega t}$.

Starting from Eq. \eqref{Eq:eqvideal_appendix}, one can  derive Eq. \eqref{Eq:eqvfinal} in the case of a polytropic equation of state for which Eq. (\ref{c2}) holds. 
By considering, moreover, Eq. (\ref{GD}) at constant temperature, Eq. (\ref{LDAplanar}) and Eq. \eqref{Eq:gamma}, one can finally rewrite:
\begin{multline}
\label{Eq:first}
\nabla \left[\left(\frac{\partial P}{\partial n}\right)_{\bar{s}}\left(\nabla \cdot {\bf v}\right)\right] = \left(\frac{\partial P}{\partial n}\right)_{\bar{s}} \nabla \left(\nabla \cdot {\bf v}\right) \\ + q \left(\nabla \cdot {\bf v}\right) \nabla V_{ext} \left(\frac{\gamma}{q} - 1\right)
\end{multline}
which plugged into Eq. \eqref{Eq:eqvideal_appendix} yields Eq. \eqref{Eq:eqvfinal}.

\begin{acknowledgments}

The authors would like to acknowledge fruitful discussions with L. P. Pitaevskii. G. De Rosi wishes to thank also P.-É. Larré, S. Giorgini, F. Dalfovo for insightful discussions. This work has been supported by ERC through the QGBE grant, by the QUIC grant of the Horizon2020 FET program and by Provincia Autonoma di Trento.

\end{acknowledgments}

\end{document}